\documentclass[sigconf]{acmart}
\renewcommand\footnotetextcopyrightpermission[1]{} 
\AtBeginDocument{%
  \providecommand\BibTeX{{%
    \normalfont B\kern-0.5em{\scshape i\kern-0.25em b}\kern-0.8em\TeX}}}
 
\copyrightyear{2024} 
\acmYear{2024} 
\setcopyright{rightsretained} 
\acmConference[SIGIR '24]{Proceedings of the 47th International ACM SIGIR Conference on Research and Development in Information Retrieval}{July 14--18, 2024}{Washington, DC, USA}
\acmBooktitle{Proceedings of the 47th International ACM SIGIR Conference on Research and Development in Information Retrieval (SIGIR '24), July 14--18, 2024, Washington, DC, USA}
\acmDOI{10.1145/3626772.3657966}
\acmISBN{979-8-4007-0431-4/24/07}

\makeatletter \gdef\@copyrightpermission{
\begin{minipage}{0.3\columnwidth} \href{https://creativecommons.org/licenses/by/4.0/}{\includegraphics[width=0.90\textwidth]{4ACM-CC-by-88x31.eps}} \end{minipage}\hfill \begin{minipage}{0.7\columnwidth}
\href{https://creativecommons.org/licenses/by/4.0/}{This work is licensed under a Creative Commons Attribution International 4.0 License.}
\end{minipage}
\vspace{5pt} }
\makeatother

\acmSubmissionID{sp7535}

\begin{document}
\fancyhead{}
\title{Multi-Layer Ranking with Large Language Models for News Source Recommendation}

\author{Wenjia Zhang}
\orcid{0009-0004-5645-9926}
\affiliation{%
  \department{Computer Science}
  \institution{University of Warwick}
  \city{Coventry}
  \country{UK}}
\email{wenjia.zhang@warwick.ac.uk}

\author{Lin Gui}
\orcid{0000-0002-8054-9524}
\affiliation{%
  \department{Informatics}
  \institution{King's College London}
  \city{London}
  \country{UK}
}
\email{lin.1.gui@kcl.ac.uk}

\author{Rob Procter}
\orcid{0000-0001-8059-5224}
\affiliation{%
  \department{Computer Science}
  \institution{University of Warwick}
  \city{Coventry}
  \country{UK}
}
\affiliation{%
  \institution{The Alan Turing Institute}
  \city{London}
  \country{UK}
}
\email{rob.procter@warwick.ac.uk}

\author{Yulan He}
\orcid{0000-0003-3948-5845}
\affiliation{%
  \department{Informatics}
  \institution{King's College London}
  \city{London}
  \country{UK}
}
\affiliation{%
  \institution{The Alan Turing Institute}
  \city{London}
  \country{UK}
}
\email{yulan.he@kcl.ac.uk}


\begin{abstract}

To seek reliable information sources for news events, we introduce a novel task of expert recommendation, which aims to identify trustworthy sources based on their previously quoted statements. To achieve this, we built a novel dataset, called NewsQuote, consisting of 23,571 quote-speaker pairs sourced from a collection of news articles.
We formulate the recommendation task
as the retrieval of experts based on their likelihood of being associated with a given query. We also propose a multi-layer ranking framework employing Large Language Models to improve the recommendation performance. Our results show that employing an in-context learning based LLM ranker and a  multi-layer ranking-based filter significantly improve both the predictive quality and behavioural quality of the recommender system.

\end{abstract}







\maketitle

\section{Introduction}

Identifying credible information sources plays a crucial role in ensuring the integrity and accuracy of journalism. First, the reliability of the source of a claim strengthens the groundwork for accurate assessments regarding the authenticity of the claim \cite{chu2012detecting,mukherjee2013spotting,shu2019role,al2020lies,qureshi2022deception}. Second, external information sources can provide additional evidence for verifying claims' veracity  \cite{vladika2023scientific,thorne-etal-2018-fact,schuster-etal-2021-get,augenstein-etal-2019-multifc,hanselowski-etal-2019-richly,kotonya-toni-2020-explainable-automated,wadden-etal-2020-fact,saakyan2021covid,karadzhov-etal-2017-fully}. Most fact-checking research assumes that the information provided by a reputable platform is trustworthy. However, it is crucial to recognise that these platforms can only be considered as secondary sources comparison to the primary speakers quoted in articles. In practice, it is essential for fact-checkers and journalists to seek out and consult credible primary sources within the relevant field to gain insights \cite{procter2023some}.

Tailored software toolkits are designed to accommodate various strategies for evaluating information sources, including the consideration of diversity \cite{shang2022dianes} and topic relevance \cite{cao2019belink}.  
Moreover, datasets have been curated to facilitate insights into the extraction and attribution of quotes in news articles \cite{
vaucher2021quotebank,zhang2021directquote}. 
Quote Erat is the first interface designed for interactive exploration of large-scale corpora of quotes from the news domain \cite{vukovic2022quote}, allowing users to retrieve relevant quotes and articles based on their queries. 
These systems do not directly recommend potential information sources, such as experts or organizations, in response to user queries. 

In the booming of Large Language Models (LLMs) \cite{DBLP:conf/nips/BrownMRSKDNSSAA20}, content-based recommendation methods have experienced a surge in effectiveness, particularly due to the advancements in In-Context Learning\cite{DBLP:journals/corr/abs-2305-14973}. Various prompting techniques have been designed to enhance the capabilities of recommender systems, serving diverse roles such as being a recommender \cite{dai2023uncovering,liu2023chatgpt}, a ranker \cite{hou2023large}, or a designer \cite{lyu2023llm,wang2023generative}. In these approaches, LLMs function as a substitute for a specific step within recommendation systems.


In this paper, we present an expert-based recommendation system for identifying information sources by analyzing quote-speaker pairs within news articles. Our contributions are: 
(1) We describe the construction of a novel dataset, named NewsQuote\footnote{The dataset can be accessed at \url{https://zenodo.org/records/11032190}}, which contains quote-speaker pairs extracted from a collection of news articles. 
(2) We implement a source recommendation system that retrieves potential information sources for a given query by leveraging historical quotations.
(3) We enhance the expert retrieval approach by integrating a multi-layer ranking-based filtering mechanism consisting of LLM rankers.
(4) Experimental results show that multi-layer LLM ranking improves the predictive quality of the recommendation system while mitigating  popularity bias.



\section{Data Construction}

We built our \textsc{NewsQuote} dataset from the AYLIEN coronavirus dataset\footnote{\url{https://aylien.com/resources/datasets/coronavirus-dataset}}, which contains news articales published between November 2019 and August 2020. 
Apart from text, each article is also accompanied with the meta data such as authors, 
keywords, 
summary, source, publishing time, 
topical categories coded by both the Interactive Advertising Bureau (IAB) taxonomy\footnote{\url{https://www.iab.com}} and the IPTC NewsCodes\footnote{\url{https://iptc.org/standards/newscodes/}}, as well as the recognized entities and entity links from the DBpedia.

\noindent\textbf{Data Deduplication}
As the same news story may be posted by multiple sources, we removed news articles that are similar to what have already been published. 
News articles were first sorted in chronological order. News duplicates were then detected using a RoBERTa classifier\footnote{\url{https://huggingface.co/vslaykovsky/roberta-news-duplicates}} trained with title-body pairs using semi-supervised learning \cite{ruckle-etal-2019-neural}. For processing efficiency, the dataset was split into 16 equal-sized subsets. For each subset, the titles and the first sentence of the news summaries of the temporally-ordered news articles were sequentially fed as input to the RoBERTa classifier. Any duplicates were removed. 
After data deduplication, 158,325 news articles were kept.

\begin{table}[H]
\centering
\footnotesize
\resizebox{0.95\columnwidth}{!}{
\begin{tabular}{l r r r}
\toprule
 & \textbf{Test} & \textbf{Valid} & \textbf{Train} \\ 
\midrule
\textbf{No. of samples} & 2000 & 1961  &  19610\\
\textbf{No. of articles} & 1749 & 1741 & 13728 \\
\textbf{No. of speakers} & 936 & 923 & 2843 \\
\textbf{Avg. quote length} & 18.24 & 16.04 & 16.12 \\
\textbf{No. of domains} & 184 & 185 & 258 \\
\textbf{No. of news categories} & 441 & 428 & 634 \\
\textbf{No. of keywords} & 21672 & 21769 & 70560 \\
\bottomrule
\end{tabular}}
\caption{The \textsc{NewsQuote} Dataset statistics. Speaker is the information source, domain is the platform where article was published. Quote length is the No. of words.}
\label{tab:QAresults}
\end{table}
\vspace{-2em}

\noindent\textbf{Quote Trigger Word Filtering}
For each of the selected articles, we segmented the the main body into sentences, and then used a pre-trained BERT-based semantic role labeling model \cite{shi2019simple} to extract verbs (or predicates), subjects, and objects. We obtained a candidate verb list sorted by their occurrence frequencies. After manually checking the most frequent candidate verbs with occurrences over 100, we identified 352 quote trigger words which are more likely indicative of direct or indirect quotes. 
Some of the verbs are clearly indicative of quotes, such as `\emph{said}', while others may not be associated with quotes in a traditional sense, for example, `\emph{tweet}'. After identifying the quote trigger words, we only kept the sentences with at least one trigger word, one subject, and one object. The subject is regarded as a potential speaker and the object is considered as a potential quotation. To ensure that the quotations are informative, we also require that the length of the object should be more than three words.

\noindent\textbf{Speaker and Quote Filtering}
We required that the subject of a candidate sentence should be a person or an organisation, and therefore identified potential speaker entities via the accompanied DBpedia ontology labels\footnote{\url{http://mappings.dbpedia.org/server/ontology/classes/}} in the dataset. 
As the same subject could have multiple mentions, we use the DBPedia entity links for entity resolution and normalisation. In addition, 
we required a named entity to appear at least twice in the dataset. Finally, to avoid the sentence split error, we required the quotation marks to be paired in sentences that contain direct quotes and mixed quotes. 

\noindent\textbf{Test Set Annotation}
Since in practice, given a topic, we can only identify experts based on their previous quotes published in earlier news articles, 
we divide the dataset into training, validation and test sets by 
the publishing timestamps of news articles, ensuring speaker-quote pairs in the validation and the test sets occurred later than those in the training set. 
To ensure data quality, samples in the test set were manually screened 
by one annotator. 



\noindent\textbf{Dataset Statistics}
Our data covers three categories of quotes: direct quote, indirect quote and mixed quote.
In short, direct quotations are placed inside quotation marks while indirect quotations are not, and a mix of direct and indirect quotations have only part of the quotations placed inside quotation marks. We roughly estimated the weight of each quotation type on the dataset by the number and position of quotation marks: 81\% for indirect quotes, 11\% for direct quotes, and 7\% for mixed quotes. 
In the test set, there are 1,582 (79\%) indirect quotes, 178 (9\%) mixed quotes, and 240 direct quotes (12\%).  
Table \ref{tab:QAresults} shows the statistics of our final \textsc{NewsQuote} dataset. In summary, we have a total of 23,571 English speaker-quote pairs with 2,843 speakers from 263 global domains. 


\begin{figure*}[h]
    \centering
    \includegraphics[width=0.95 \textwidth]{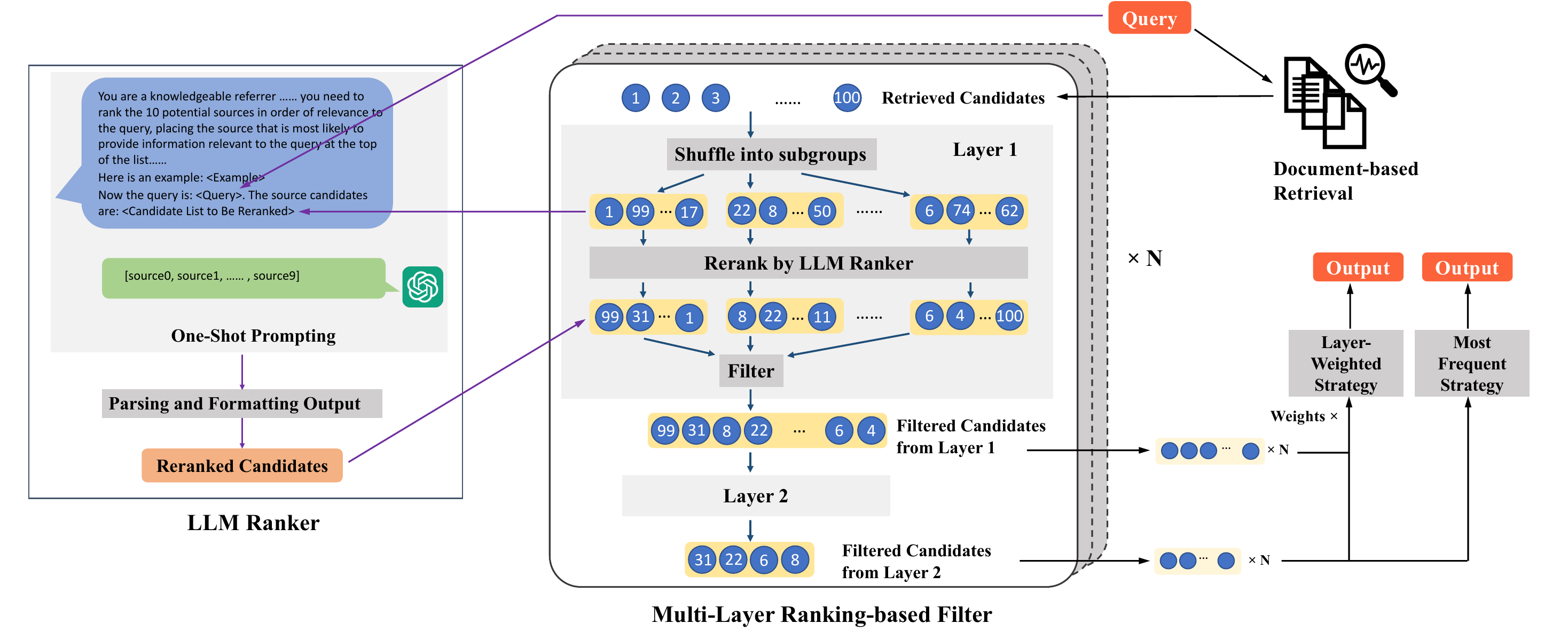}
    \vspace{-1em}
    \caption{Illustration of the LLM Ranker and the Multi-layer Ranking-based Filter}
    \label{fig:framework}
    \vspace{-1em}
\end{figure*}

\section{Approaches}
In our dataset, each sample $S_{i}$ consists of a context  $d_{i}$, a quote-speaker pair $(q,e)_{i}$ extracted from the context, and metadata $m_{i}$ inherited from the article where it originates. The context contains three sentences, the primary sentence containing the speaker and quote, its preceding sentence and the following sentence. The metadata are defined at the document level and are sourced from the AYLIEN coronavirus dataset. 
We formulate source recommendation as a retrieval problem, aiming at identifying sources capable of commenting on the topic discussed in a given query, guided by the historical quotes of the sources in news reports, and subsequently ranking them by their relevance to the query. In our scenario, each context is treated as a document, and a preprocessed news article title serves as the query.


\subsection{Expert Finding} 

We implemented both candidate-based and document-based expert retrieval methods \cite{balog2009language} for this source recommendation task. 

The \textit{Candidate-based Expert Retrieval} assumes that each term in the query is sampled identically and independently, also that the document and the expert source candidate are conditionally independent. The candidate-based approach estimates $P(k|e)$ by: 

\begin{footnotesize}
\begin{gather}
P(k|e)=\prod_{t\in k} \{ (1-\lambda)(\sum_{d\in D}p(t|d)p(d|e))+\lambda p(t) \} ^{n(t,k)},\nonumber \\
\lambda=\frac{\beta}{\beta+n(e)},  \quad
\beta=\frac{\sum_{E}|\{d: n(e,d)>0\}|\cdot|d|}{|E|}, \nonumber
\end{gather}
\end{footnotesize}
\noindent where $\lambda$ is the smoothing parameter, $p(t|d)$, $p(d|e)$ and $p(t)$ are the conditional probability of a term $t$ in document $d$, the conditional probability of a document $d$ given source $e$, and the probability of term $t$, respectively. Both $p(t|d)$ and $p(t)$ are estimated by maximum likelihood. 
$n(t,k)$ is the number of times a term $t$ appears in the query $k$, $n(e,d)$ is the occurrence frequency of an source $e$ appeared in the document $d$, $n(e)$ is the total number of tokens in all documents associated with the source $e$, $|d|$ is the average document length, and $|E|$ is the total number of sources.

The \textit{Document-based Expert Retrieval} approach searches for sources via relevant document collection by assuming the conditional independence between the query and candidate, and estimates the probability of a term $t$ in each document. The $P(k|e)$ is then calculated as: 

\begin{footnotesize}
\begin{gather}
\small
  P(k|e)=\sum_{d\in D} \{\prod_{t\in k}((1-\lambda)p(t|d)+\lambda p(t))^{n(t,k)}\}p(d|e), \nonumber \\
\lambda=\frac{\beta}{\beta+n(d)},  \quad
\beta=|d|, \nonumber
\end{gather}
\end{footnotesize}
\noindent where $n(d)$ is the length of document $d$.

In both expert finding approaches, the document-candidate associations $p(d|e)$ is estimated by a simple Boolean model.

\subsection{LLM-based Multi-Layer Ranking}
Purely probability-based approaches face significant constraint due to the limitation of the document corpus, making it challenging to efficiently enhance models with additional knowledge in advance, especially when facing unpredictable news topics. Instead, we take advantage of Large Language Models to improve the performance of the recommender system.

\noindent\textbf{LLM Ranker}
After obtaining a list of recommendations returned by the Expert Finding methods, we employ LLMs to rerank the top recommended candidates. This is achieved by devising a one-shot prompt template:

\leftskip 10pt
\rightskip 10pt
  \textit{"You are a knowledgeable referrer.
  Given a query and the 10 potential information sources (which may include both individuals and organizations) retrieved based on the query, you need to rank the 10 potential sources in order of relevance to the query, placing the source that is most likely to provide information relevant to the query at the top of the list.
  Return the new rank of sources only in the form of python list (exactly the same form of the given list, just rerank it), please do not provide other words except for the list. 
  Here is an example:
  Query: <Example Query>.
  10 potential sources are: <Candidate List for the Example Query>, and then the output should be:  <Reranked Candidate List>.
  Now the query is: <Query>. The source candidates are: <Candidate List to Be Reranked>"}

\leftskip 0pt
\rightskip 0pt

\noindent\textbf{Multi-layer Ranking-based Filter}
When using an LLM ranker to improve ranking quality, the system's predictive efficacy still depends on the probability-based retrieval approach that provides the initial candidates. To expand the pool of candidates considered for rankings, we propose a \textit{Multi-layer Ranking-based Filter}. This filter uses the one-shot LLM ranker as a component to generate result from a wider range of candidates by sequentially sorting and filtering through multiple layers. As shown in Fig. \ref{fig:framework}, in each layer, the input candidate list is first randomly divided into several subgroups, and then each subgroup is fed into an LLM ranker. The filter keeps only a subset of the top-ranked candidates from each reranked subgroup candidate list, and concatenates them into a new list of candidates for the subsequent layer. Inspired by the Sequential Monte Carlo Steering \cite{lew2023sequential}, the LLM-based filter operates N times to clone promising candidates and cull low-likelihood ones.

\noindent\textbf{Filter Strategy}\label{Filter Strategy} In our framework, the aforementioned filter undergoes multiple iterations to generate confidence scores for the resulting candidates.  We employ two strategies to determine the final output based on the frequency of retention by the filter. The \textit{Most Frequent Strategy} only considers the output from the last layer of the Multi-layer Ranking-based Filter. It selects  candidates with the highest frequency of occurrence across the repeated runs. On the other hand, the \textit{Layer-Weighted Strategy} considers the outputs of all filter layers. It calculates the average repetition frequency of each layer using a pre-defined set of weights, and then returns candidates with the highest weighted occurrence score.

\section{Experiments}
In this section, we describe how we set up experiments and evaluate the  results.

\subsection{Setup}
In experiments, we utilized context samples from the training set as the document database and used article titles from the test set as queries. There could be more than one source of information corresponding to an article title query. We employed the GPT-3.5-Turbo and GPT4-Turbo as LLM rankers, and GPT-3.5-Turbo as the Multi-layer Ranking-based Filter. The filter comprised a 2-layer structure: (1) In the first layer, we initialized with 100 candidates returned by the document-centric approach, dividing them randomly into 10 groups. Each group then underwent reranking by an LLM ranker, with the top 5 candidates from each group being retained; (2) Subsequently, in the second layer, the top 50 candidates from the first layer were shuffled, divided into 5 groups, and reranked by an LLM ranker. Only the top 2 candidates from each group, as returned by the LLM ranker, were kept. Upon each run of the 2-layer ranking-based filter, we obtained a list of 10 candidates. After repeating this process 20 times, we determined the final output by adopting different filter strategies mentioned in \ref{Filter Strategy}, resulting in the top 20 recommendations. 

\subsection{Metrics}

\noindent\textbf{Recall} is the proportion of correctly identified information sources in the top recommendations out of the total number of sources quoted in the article.

\noindent\textbf{Mean Averaged Precision (MAP)} is the mean of the precision at the points where the relevant sources were retrieved.

\noindent\textbf{Normalized Discounted Cumulative Gain at K}
first discounts the gain scale at the $i$-th rank position by $\frac{1}{\log_{2}(i)}$, then adds up the converted gain scales up to rank $k$, and finally normalizes the result by the ideal ranking order.

\noindent\textbf{Diversity} is calculated as the average pairwise distance between items appeared in the recommendation list. To encode entity features, we trained Wikipedia2Vec embeddings \cite{yamada2018wikipedia2vec} on a Wikipedia database publish in Jan 2024.

\noindent\textbf{Coverage} shows the percentage of information sources from the training dataset being recommended on the test set.

\noindent\textbf{Average Recommendation Popularity} is the averaged popularity of the recommended items on the training set. Here, we use the number of occurrences in the training set as the popularity of the information source.

\section{Results}

\begin{table}[H]
\footnotesize
\centering
\resizebox{0.95\columnwidth}{!}{

\begin{tabular}{l | c c c c }
\toprule
  & CER & DER & ReDE4& ReDE3.5 \\
\midrule
Recall &  0.2454 & 0.3241  & 0.3244  & \textbf{0.3249} \\
MAP  &  0.1342 & 0.2016  & \textbf{0.2257}  & 0.1822 \\
NDCG10 & 0.1619  &  0.2327 & \textbf{0.2512}  & 0.2176 \\

\bottomrule
\end{tabular}}
\caption{Results of the Expert Retrieval baselines and LLM ranker. CER is the Candidate-based Expert Retrieval, and DER is the Document-based Expert Retrieval. ReDE3.5 used GPT-3.5 to rerank the top 10 candidates returned by the DER, while ReDE4 used GPT-4. }
\label{tab:results1}
\end{table}
\vspace{0em}
As baselines, the Document-based Retrieval method performed much better than the Candidate-based Retrieval method, scoring 7\% to 8\% higher in Recall, MAP, and NDCG@10. This disparity is likely attributed to the fact that the training set is organized by documents rather than by information source candidates. When we applied the LLM ranker on the top 10 candidated returned by the Document-based Retrieval approach, GPT-4 improved the MAP and NDCG@10 by 2\%, but GPT-3.5 did not yield improvement. This may be because GPT-4 possesses more up-to-date knowledge compared to GPT-3.5. Interestingly, both the GPT-4 and GPT-3.5 rankers yielded a slight improvement in Recall, albeit negligible. We attribute this phenomenon to the stochastic nature of LLM generation. 

Although the single GPT-3.5 ranker did not perform well in reordering the top 10 candidates, integrating a multi-layer structural filter into the baseline approach resulted in a modest improvement in Recall of the top 20 candidates from the top 100 candidates by 0.5\% to 2\%. Given the task of  recommending sources for news stories, the diversity and comprehensiveness of sources are also important alongside their reliability in providing accurate information. Following the application of the Multi-layer Ranking-based Filter, the average recommendation popularity decreased significantly from 71.68 to around 40. This indicates that our approach has the capability to divert the baseline from favoring popular sources. Moreover, allocating more weights from the last layer to the first layer led to a slight decrease in recall, while increasing MAP, NDCG@10, and popularity bias. Consequently, we conclude that the multi-layer structure sacrifices ranking precision to increase recall and mitigate popularity bias.

\begin{table}[H]
\footnotesize
\centering
\resizebox{0.98\columnwidth}{!}{
\begin{tabular}{l | c c c c c }
\toprule
  & DER & MRF & MRF0& MRF1&MRF2\\ 
\midrule
Recall &  0.3909& \textbf{0.4098}  & 0.4080 & 0.4012 & 0.3940\\
MAP & \textbf{0.2056} & 0.1719 & 0.1808 & 0.1805 & 0.1795\\
NDCG10& \textbf{0.2318} & 0.2059 & 0.2126 & 0.2127 & 0.2117 \\
Diversity & 10.34 & \textbf{5.889} & 5.896 & 5.894 & 5.894 \\ 
Coverage & 0.2085 & \textbf{0.2595} & 0.2562 & 0.2497 & 0.2497 \\ 
ARP & 71.68 & \textbf{39.11} & 39.44 & 39.63 & 40.05 \\ 
\bottomrule
\end{tabular}}
\caption{Evaluations of the Multi-layer Ranking-based Filter  on the top 20 candidates. DER denotes the Document-based Expert Retrieval. MRF represents the DER results augmented by the multi-layer ranking-based filter using the most frequent strategy. MRF0(MRF1, MRF2) denotes the DER results augmented by the multi-layer ranking-based filter with the layer-weighted strategy and layer weights [0.25,0.75]([0.5,0.5],[0.75,0.25]).}
\label{tab:results2}
\end{table}
\vspace{0em}

In evaluating the results, the golden answers refer to sources that were actually cited in news reports, averaging only 1.052 golden answers per query. These golden answers received a score of 1, while all other candidates received a score of 0 when calculating Recall, MAP, and NDCG@10. 
As such, we posit that the potential of the recommender system could be underestimated. In reality, sources not cited in  news articles may still hold value, and therefore should be scored higher than zero. 

\section{Conclusion and Future Work}

This paper outlines the development of a novel dataset comprising quote-source pairs extracted from news articles. It also presents a recommendation task, aimed at identifying information sources for a given news topic query. Based on the \textsc{NewsQuote} dataset, we explored the Expert Finding baselines for this task, and proposed a Multi-layer Ranking-based Filter, which integrates LLM rankers to enhance the capability of the recommender system. Results show that the Multi-layer LLM Ranker improves predictive accuracy and mitigates popularity bias, albeit at the expense of ranking precision. In future work, we will engage with a more diverse document corpus and utilize human evaluations to provide more practical ratings.


\section*{Acknowledgements}
This work was supported in part by the UK Engineering and Physical Sciences Research Council (EP/V048597/1, EP/X019063/1) and a Turing AI Fellowship (EP/V020579/1, EP/V020579/2). 

\bibliographystyle{ACM-Reference-Format}
\bibliography{articlebib}


\end{document}